\documentclass[11pt]{article}

\let\ssection=\section
\renewcommand{\section}{\setcounter{equation}{0}\ssection}
\usepackage{bm}
\usepackage{amsmath,dsfont}

\usepackage{graphics,amssymb,amsmath,amsthm,amscd,array}

\def\parag{\hfil\break} 
\def\kikezd{\parag\underbar}

\def\p{{\partial}}

\def\bLambda{{{\bf \Lambda}}}

\def\su{{\rm su}}
\def\SU{{\rm SU}}

\def\SO{{\rm SO}}

\def\ort{{\rm o}}
\def\smallover#1/#2{\hbox{$\textstyle\frac{#1}{#2}$}}
\def\2{{\smallover1/2}}

\def\dAlembert{\vcenter {
    \hbox {\vrule height8pt width0.4pt depth0.0pt
           \vrule height8pt width7.2pt depth-7.6pt
           \vrule height8pt width0.4pt depth0.0pt
           \kern-8pt
           \vrule height0.4pt width8pt depth0.0pt
          \,}}}

\def\I1{{\mathds{1}}}
\def\D{{D\mkern-2mu\llap{{\raise+0.5pt\hbox{\big/}}}\mkern+2mu}}

\def\ri{{\mathrm i}}

\title{\bf Applications of chiral supersymmetry for spin fields in self-dual backgrounds}

\author{L. Feh\'er\footnote{On lave from Bolyai Institute, University of Szeged. Present address:
Research Institute for Particle and Nuclear Physics, Budapest,
Hungary. e-mail: lfeher-at-rmki.kfki.hu}
 \quad P. A. Horv\'athy\footnote{Department de Math\'ematiques  Universit\'e d'Avignon.
 Present address: Laboratoire de Math\'ematiques et de
Physique Th\'eorique, Universit\'e de Tours, France. e-mail:
horvathy-at-lmpt.univ-tours.fr}
\quad and \quad
L. O'Raifeartaigh
\\[16pt]
Dublin Institute for Advanced Studies, Ireland}

\begin{document}

\maketitle

\begin{abstract}
Due to chiral supersymmetry the (nonzero mode) spectral and symmetry
properties of a $4$-dimensional, self-dual Dirac-Yang-Mills operator $\D$
can be recovered from those of the corresponding scalar Laplacian $D^2$. It is shown that a similar result holds for higher spins, and that in the $4$-vector
case the supersymmetric partners are $-D^2\I1_4$ and the fluctuation operator. The reduction to $D^2$ is used to simplify previous analyses
of the  (nonzero mode) spectrum of $\D$ and of the fluctuations for a BPS-monopole, and to explain the Kepler and
$\su(2/2)$ (super)symmetries of a system studied recently by D'Hoker and  Vinet.
\end{abstract}
\bigskip\noindent
Published as {\sl Int. Journ. Mod. Phys.}  {\bf A4}, 5277-5285 (1989).
\bigskip

\goodbreak
\section{Introduction}
It is known that the Dirac-Yang-Mills (DYM) operator $\D=\gamma^\mu D_\mu$, $D_\mu=\p_\mu-\ri A_\mu$,
is chiral-supersymmetric in any even dimensions \cite{1}, and it is easy to verify that for
 a $4$-dimensional, self-dual gauge potential $A_\mu$, the squared operator $\D^2$ reduces to the scalar Laplacian $D^2$ (times the unit matrix) on one of the supersymmetric sectors. The reduction of $\D^2$ to $D^2$ has been used in constructing propagators in instanton  background \cite{2}.
The purpose of the present paper is to show how the reduction can be used to simplify and clarify some recent work on the discrete spectra and symmetries of self-dual DYM operators. We first show how the discrete
eigenvector system (apart from the zero modes) and the symmetries of $\D$ may be reconstructed from
those of $D^2$, and then use $D^2$ to

(i) simplify and generalize to any isospin the previous analyses \cite{3,4} of the (nonzero mode)
spectrum of the DYM operator of a Bogomolny-Prasad-Sommerfield (BPS) monopole \cite{5}
(the zero-modes have been completely analyzed in Refs. \cite{6} and \cite{7}),

(ii) explain the remarkable but complicated `dynamical' Kepler and supersymmetries
of the self-dual $\D^2$ found recently by D'Hoker and Vinet \cite{8},

(iii) show how the DYM supersymmetry has an immediate generalization
from Dirac-spinor ($m=0$) to $D^{(1/2,m)}\oplus D^{(m,1/2)}$
wave functions for any $m=0,\2,1,\smallover3/2,\dots$ (note that
$D^{(p,q)}$ denotes the representation of ${\rm SO}(4)
\simeq\SU(2)\times\SU(2)$, which is obtained by taking the tensor product of the angular momentum $p$ and $q$ representations of the
left and right $\SU(2)$ subgroups, respectively), and that in the $4$-vector case ($m=\2$) the SUSY partners are $-D^2\I1_4$ and the operator on the left-hand side of the fluctuation equation \cite{9}
\begin{equation}
\big[-D^2\delta_{\mu\nu}+2\ri F_{\mu\nu}\big]\delta A^\nu=
\omega^2(\delta A^\mu),
\qquad
F_{\mu\nu}\equiv \ri [D_\mu,D_\nu].
\label{(1.1)}
\end{equation}
The latter result throws a new light on the  well-known fact \cite{10,9,3,4} that the $\omega\neq0$ eigenfluctuations can be expressed in terms of the eigenfunctions of either $\D$ or $D^2$.

The starting point is the usual $4$-dimensional Euclidean DYM operator
in the chiral (supersymmetric) basis, namely,
\begin{equation}
Q=\left[\begin{array}{cc}
0&S\\[6pt]
S^\dagger&0
\end{array}\right]\equiv \ri\D=
\left[\begin{array}{cc}
0&\sigma^kD_k-\ri D_4\\[6pt]
-\sigma^kD_k-\ri D_4&0
\end{array}\right],
\qquad
\gamma_5=\left[\begin{array}{cc}
\I1_2&0\\[6pt]0&-\I1_2
\end{array}\right],
\label{(1.2)}
\end{equation}
from which one has, for the squared operator,
\begin{equation}
Q^2=\left[\begin{array}{cc}
SS^\dagger&0\\[6pt]
0&S^\dagger S
\end{array}\right]
=-\D^2=
\left[\begin{array}{cc}
-D^2-\sigma^k(B_k+E_k)&0\\[6pt]
0&-D^2-\sigma^k(B_k-E_k)
\end{array}\right],
\label{(1.3)}
\end{equation}
the $\sigma$'s being the three Pauli matrices. From (\ref{(1.3)}) it is evident that for self-dual gauge fields
($E_k=B_k$) the operator $\D^2$ reduces to $D^2$ on one of the supersymmetric sectors.

First we wish to show that, at least for the discrete spectrum, the eigenvector system and the symmetries of $\D$ can be reconstructed from those of $D^2$. For this one
notes that if, for a general QMSUSY system, the zero eigenmodes of $Q$ are excluded from the Hilbert space, the operator
\begin{equation}
U=S\,\frac{1}{\sqrt{S^\dagger S}}
\label{(1.4)}
\end{equation}
becomes unitary (at least on the bound state sector we are interested in), and intertwines the supersymmetric partners of $Q^2$, i.e.,
\begin{equation}
SS^\dagger=U(S^\dagger S)U^\dagger.
\label{1.5)}
\end{equation}
It follows that if $\psi$ is an eigenvector of $S^\dagger S$ with
eigenvalue $\omega^2>0$, then $U\psi$ is an eigenvector of $SS^\dagger$ with the same eigenvalue and
\begin{equation}
\psi_{\pm\omega}\equiv\frac{1}{\sqrt{2}}\left(
\begin{array}{c}
U\psi\\[6pt]\pm\psi\end{array}\right)
\label{(1.6)}
\end{equation}
are eigenvectors of $Q$ with eigenvalues $\pm\omega$
($\omega>0$), respectively. Similarly, if
$G$ is a symmetry group of $S^\dagger S$ then $UGU^\dagger$ is a symmetry group
of $SS^\dagger$ and the diagonal subgroup of $(UGU^\dagger)\times G$
is a symmetry group of $Q$. Applying these results to
the case $Q=\ri\D$ for a self-dual $A_\mu$, one sees at once that the discrete eigenvector system and the symmetries of $\D$ can be reconstructed from
those of $D^2$. Examples of the symmetry reconstruction will be given in Sec. 3.

\section{Nonzero mode spectrum for the BPS monopole}

Let us first apply the reduction of $\D^2$ to $D^2$ to the study of the bound state (nonzero mode) spectrum of $\D$ for the gauge field of an $\SU(2)$ (isospin),
static, spherically symmetric BPS-monopole. This is the
special case in which, in the usual three-dimensional notation, the self-dual gauge potential
$A_\mu$ is
\begin{equation}\begin{array}{lll}
A_4^a=\Phi^a=-\displaystyle\frac{x^a}{r} \displaystyle\frac{H}{r},&\qquad
&A_i^a=\epsilon_{aik} \displaystyle\frac{x^k}{r^2}\,(1-K),
\\[12pt]
H=r\coth r-1,
&&K= \displaystyle\frac{r}{\sinh r},
\end{array}
\label{(2.1)}
\end{equation}
($A_4^a$ being identified physically as a Higgs field
$\Phi^a$,
and correspondingly $E_k\to D_k\Phi$).
The bound state problem in this background has been extensively investigated before \cite{3}
for an isospin $1$ particle on the basis of the Jackiw-Rebbi \cite{6} angular momentum
analysis of $\D$ (see also ref. \cite{4}). If we exclude the zero modes then, according to the
arguments given above, it is sufficient to study the scalar Laplacian $D^2$, and for
the gauge potential (\ref{(2.1)}), this Laplacian (for $x^4$-\textit{independent} wave-functions) is easily computed to be
\begin{equation}
-D^2=-\bigtriangleup_r+\frac{1}{r^2}\Big\{\Lambda^2
+K(K-2){\mathbf T}^2-2KI_0+\big[H^2-(1-K)^2\big]
Q_{em}^2\Big\},
\label{(2.2)}
\end{equation}
where $\bigtriangleup_r$ is the usual radial Laplacian, $\Lambda^i$ is the total angular momentum,
\begin{equation}
\Lambda^i=L^i+T^i,
\label{(2.3)}
\end{equation}
$L^i$ and $T^i$ are the orbital angular momentum and isospin
$\su(2)$ generators, respectively, and $I_0$ and $Q_{em}$ are the
isospin-orbit coupling and the electric charge operators defined as
\begin{equation}
I_0=L_k\,T^k,
\qquad\hbox{and}\qquad
Q_{em}=-\frac{x^a}{r}\,T_a
\quad\left(=\frac{\Phi^a}{|\Phi|}\,T_a\right).
\label{(2.4)}
\end{equation}
The operator $D^2$ acts on the function space
${\cal H}=L^2(rdr)\oplus L^2(S^2)\oplus C^{2t+1}$ and since it commutes with the total
isospin ${\bf T}^2=\sum_i (T^i)^2$ and the total angular momentum
vector $\Lambda^k$, it can be studied for each set of eigenvalues
$\big\{t(t+1), \lambda(\lambda+1),\lambda_3\big\}$ of the
operators $\big\{{\mathbf T}^2, \bLambda^2,\Lambda_3\big\}$ separately.
This labeling of the nonradial part of ${\cal H}$ leaves a
$(2t+1)$-dimensional degeneracy, which can be removed in a natural
way by labeling with the $(2t+1)$-eigenvalues $-t\leq q\leq t$ of $Q_{em}$.
Since $Q_{em}$ does not commute with $D^2$ the various $Q_{em}$ eigenstates are
coupled, but because $D^2$
(and $\Lambda^i$) commute with the space-parity operator $P$, it is possible to
decompose the $(2t+1)$-states into two decoupled sets of opposite parity.
Because $P$ and $Q_{em}$ anticommute, the eigenstates
of $P$ are simply the sums and differences of the eigenstates of $Q_{em}$ for
fixed $q^2$ and the subspaces of definite parity are  (in general) of dimension $(t+\2)$
for $t=$ half-odd-integer, and $t$ and $(t+1)$ for $t=$ integer, the $(t+1)$-dimensional
sector containing the electrically neutral state ($Q_{em}$=0). Since the angular operators
$I_0$ and $Q_{em}^2$ which occur in $D^2$ do not commute, and are irreducible on the
parity eigenspaces for fixed $(t,\lambda,\lambda_3)$, the radial equations within
each parity sector cannot be further decoupled.

For example, for the case $t=1$, to which the problem of nonzero mode fluctuations around the monopole can be reduced (see Sec. 4), one sees that for each value of $\lambda,\lambda_3$ ($\lambda=1,2,\dots$) there are one uncoupled and two coupled radial equations, and that the operator $-D^2$ takes the form
\begin{equation}
\left[-\Delta_r+\frac{\lambda(\lambda+1)}{r^2}+
\frac{H^2+K^2-1}{r^2}\right]
\label{(2.5a)}
\end{equation}
and
\begin{equation}
\left[-\Delta_r+\frac{\lambda(\lambda+1)}{r^2}
+\frac{1}{r^2}\left(\begin{array}{cc}
2K^2&0\\[6pt]0&(H^2+K^2-1)
\end{array}\right)
-\frac{2K\sqrt{\lambda(\lambda+1)}}{r^2}
\left(\begin{array}{cc}
0&1\\[6pt]1&0
\end{array}\right)\right]
\label{(2.5b)}
\end{equation}
in the respective parity sectors. (Note that $Q_{em}^2$ is
also diagonal in (\ref{(2.5a)})-(\ref{(2.5b)})). For the lowest angular momentum $\lambda=0$ only the upper ($Q_{em}=0$) half of (\ref{(2.5b)}) survives.

The main question concerning the operator (\ref{(2.2)}) is whether
 it has a discrete spectrum, i.e., admits bound states, and since the functions $H(r)$ and $K(r)$
 are well-behaved at the origin, this question can be investigated by concentrating on the asymptotic
 ($r\to\infty$) form of $-D^2$, which is easily seen to be
\begin{equation}
-D^2\Big|_{r\to\infty}=
\left[-\Delta_r+\frac{\bLambda^2}{r^2}
-2\frac{Q_{em}}{r}+Q_{em}^2\right]\, .
\label{(2.6)}
\end{equation}

In this limit the operator $I_0$ drops out, the ($2t+1$) equations for fixed $(t,\lambda,\lambda_3)$ completely decouple, and one obtains a set of Schr\"odinger operators with Coulomb potentials, and a free Schr\"odinger operator for
$Q_{em}=0$. Thus if (\ref{(2.6)}) were the true operator, there would be a discrete Coulomb spectrum (for $0<E<Q_{em}^2$) for each nonzero value $q$ of $Q_{em}$. For the true operator (\ref{(2.2)}), one would therefore expect to have bound states in all irreducible sectors except those which contain the neutral ($Q_{em}=0$) states. Thus for $t=$ half-odd integer one would expect to have bound states in both parity sectors, and for $t=$ integer, one would expect to have them only in the $t$-dimensional sector (in which $Q_{em}^2\neq0$). Furthermore, since the true operator
(\ref{(2.2)}) differs appreciably from (\ref{(2.6)}) only in
the core region ($r\leq1$), one would expect the true bound state spectra to be close to the Coulomb spectra for all but the lowest eigenvalues of the angular momentum. These expectations have been confirmed \cite{11} numerically for $t=\2$ and $t=1$.
 Concentrating on the $t=1$ case, we note that the bound states of (\ref{(2.5a)})
 with $\lambda=1,2,\ldots$
 yield via (\ref{(1.6)}) bound states of $\ri \D$
 with $j=\lambda\pm\2=\2,\ \smallover3/2,\dots$, where the quantum number $j$ refers to
 the angular momentum operator $J^i=\Lambda^i+\sigma^i/2$ containing also the spin.
 In fact, the $j=\lambda-\2$ series of bound states constructed in this way belongs to the
 $F^{\pm}_{\,\,\mp,j}$ Jackiw-Rebbi sector \cite{6,3} and reproduce the bound states found in Ref.~[3]. The
$j=\lambda+\2$ series which is in the $F^{\pm}_{\,\,\pm,j}$ Jackiw-Rebbi sector went apparently
unnoticed in Ref.~[3].

\section{A point-like self-dual monopole}

Let us consider the case where the asymptotic form
\begin{equation}
A_4^a=\Phi^a=-\displaystyle\frac{x^a}{r}\left(1-\frac{1}{r}\right),
\qquad
A_i^a=\epsilon_{aik} \displaystyle\frac{x^k}{r^2}
\label{(3.1)}
\end{equation}
of the gauge potential (\ref{(2.1)}) is taken as the gauge potential for all $r\neq0$.
(In spite of the singularity at $r=0$ the operator $\ri\D$ is self-adjoint for (\ref{(3.1)})
just as for the Coulomb potential.) For this potential $-D^2$ is given (for all $r\neq0$) by the expression
(\ref{(2.6)}), where
\begin{equation}
\Lambda^i=\epsilon_{ijk}x^j(-\ri D^k)-Q_{em}\frac{x^i}{r}
\label{(3.2)}
\end{equation}
is now the angular momentum operator for a Wu-Yang monopole. In this case the electric charge $Q_{em}$ commutes with $D^2$ and we restrict ourselves to one of its eigensubspaces with a nonzero eigenvalue $q$. The $\Lambda^i$ then become the angular momentum of a charged particle in a Dirac monopole background, and the operator
(\ref{(2.6)}) itself becomes the Hamiltonian considered about 20 years ago by Zwanziger \cite{12} and
by McIntosh and Cisneros \cite{13} (where a term $q^2/r^2$, which comes from the Higgs field $\Phi^a$ in the above derivation, was introduced `by hand' in order to change a dyon-background potential
to a Coulomb one). In Refs.~[12] and [13] it was shown that the Hamiltonian (\ref{(2.6)}) has an
$\ort(4)/\ort(3,1)$ Kepler type symmetry, but one generated
not by the usual $H$-atom generators but by $\Lambda^i$ and the Runge-Lenz vector
\begin{equation}
K^i=-\frac{\ri}{2}\epsilon_{ijk}\big(D^j\Lambda^k-\Lambda^jD^k\big)-q^2\frac{x^i}{r}\ .
\label{(3.3)}
\end{equation}

The interesting feature of the potential (\ref{(3.1)})
is that the operator $-\D^2$ for this potential (and for fixed $q\neq 0$) is
just the $4\times 4$ supersymmetric Hamiltonian
studied by D'Hoker and Vinet \cite{8}, and one of its chiral projections is the
$g=4$ Pauli Hamiltonian studied by these authors. Once this is realized, the reason for the value $4$ of the gyromagnetic ratio, and for the symmetries of the Pauli Hamiltonian,
becomes clear: the value $4$ comes from the fact that for self-dual fields
$\sigma^k(E_k+B_k)\to2\sigma^kB_k$ in $SS^\dagger$, the `extra' $\su(2)$ is really the SUSY partner of
 the trivial spin $\su(2)$ symmetry of $S^\dagger S=-D^2\I1_2$ present in any
self-dual background, and the Kepler symmetry is simply the SUSY partner of the Kepler symmetry of the Hamiltonian (\ref{(2.6)}). In fact the symmetry generators in the lower sector are $\sigma^i/2$ and $(\Lambda^i\I1_2,K^i\I1_2)$, or equivalently $(J^i,\sigma^i/2,K^i\I1_2)$,
where $J^i=\Lambda^i\I1_2+\sigma^i/2$  is invariant with respect to the SUSY transformation $U$, and (on the subspaces $SS^\dagger=\omega^2>0$) the SUSY partners of $\sigma^i/2$ and
$K^i\I1_2$ are calculated to be
\begin{equation}
\Omega^i\equiv U\left(\2\sigma^i\right)U^\dagger
=
\frac{1}{\omega^2}\left[
\2(D^2-D_4^2)\sigma^i-D_4(\epsilon^{ijk}D_j\sigma_k)-
(\sigma^kD_k)D^i\right],
\label{(3.4)}
\end{equation}
and
\begin{equation}
U\left(K^i\I1_2\right) U^\dagger
=K^i\I1_2-\ri\epsilon_{ijk}D^j\sigma^k
+\left(\frac{q}{r}-\frac{q}{2}\right)\sigma^i
-(\sigma^kB_k)x^i+
q\Omega^i,
\label{(3.5)}
\end{equation}
and it is easy to verify that the D'Hoker-Vinet generators are linear combinations of these operators and $J^i$. Note however that what D'Hoker-Vinet call the
 `Runge-Lenz' vector is the transform of
$K^i\I1_2-q\sigma^i/2$, and not of  $K^i\I1_2$. We should like to emphasize that the operator $\Omega^i$ (\ref{(3.4)}) commutes with $SS^\dagger$ in any
self-dual background, and, moreover, the direct sum operator $\Sigma^i=\Omega^i\oplus\frac{\sigma^i}{2}$ commutes with the Dirac operator $\D$ itself. We shall
consider the D'Hoker-Vinet $\su(2/2)$ supersymmetry in Sec. 5.

\section{Fluctuations and generalized DYM supersymmetry}

Let us next consider the question of (nonzero mode) fluctuations around a self-dual background.
The fluctuation eigenmodes are \cite{9} transverse $D_\mu(\delta A_\mu)=0$ solutions
of Eq.~(\ref{(1.1)}), and what we wish to show is that, just like the DYM eigenfunctions, the
eigenfuncions  $\delta A_\mu$ of  (\ref{(1.1)}) (with $\omega^2\neq0$)
are supersymmetric partners of the eigenfunctions of the scalar Laplacian $D^2$. We shall show this by first generalizing
the DYM supersymmetry (\ref{(1.2)}), (\ref{(1.3)}) from Dirac spinors ($m=0$) to
 $D^{(1/2,m)}\oplus D^{(m,1/2)}$ wave-functions for any $m=0,\2,1,\smallover3/2,\dots$,
 then specializing the result to the $4$-vector case ($m=\2$).

The generalization is made by replacing the Dirac spinor representation space
of $\SO(4)$, $D^{(1/2,0)}\oplus D^{(0,1/2)}$, by the representation space
$D^{(1/2,m)}\oplus D^{(m,1/2)}$ and the operator $S$ in (\ref{(1.2)}) by $S: D^{(m,1/2)}\to D^{(1/2,m)}$ defined by
\begin{equation}
S^{A,k'}_{\;\; \;\;n,B'}\equiv D^\mu(\sigma_\mu)^A_{\;B'}\delta^{k'}_{\; n}\,,
\qquad
\mu=1,\dots,4,\qquad
(\sigma_4)^A_{\;B'}=-\ri\delta^A_{\;B'},
\label{(4.1)}
\end{equation}
where $A,\, B'=1,2$ and $n,k'=1,\dots,2m+1$ are the usual left and right spinor indices for $\SO(4)\simeq\SU(2)\times \SU(2)$. Since $S$ is given by a trivial matrix on the $m$-part of the representation space, the chiral projections of $Q^2$ are similar to those in Eq. (\ref{(1.3)}), namely,
\begin{eqnarray}
\big(SS^\dagger\big)^{A,k'}_{\;\; \;\; B,n'}&=&
-D^2\delta^A_{\; B}\delta^{k'}_{\: n'}-(B^i+E^i)
(\sigma_i)^A_{\; B}\delta^{k'}_{\; n'}\,,
\label{(4.2a)}
\\[8pt]
\big(S^\dagger S\big)^{k,A'}_{\;\; \;\;n,B'}&=&
-D^2\delta^{k}_{\: n}\delta^{A'}_{\; B'}-(B^i-E^i)
(\sigma_i)^{A'}_{\; B'}\delta^{k}_{\; n}\,.
\label{(4.2b)}
\end{eqnarray}
In particular, in a self-dual background the spin couplings cancel in $S^\dagger S$, so
$S^\dagger S$ becomes $-D^2\I1_{4m+2}$, i.e., it essentially reduces to the scalar
Laplacian $D^2$. (Note that the Pauli matrices play  completely different roles
 in Eqs. (\ref{(4.1)}) and (\ref{(4.2a)})-(\ref{(4.2b)}), being intertwining
 operators in (\ref{(4.1)}) and group operators in (\ref{(4.2a)})-(\ref{(4.2b)}).) For $m=\2$,
$D^{(1/2,m)}$ and $D^{(m,1/2)}$ both become the $4$-vector representations of $\SO(4)$,
and in the usual vector basis $S$ and $SS^\dagger$ take the form
\begin{eqnarray}
S_{\mu\nu}&=&-\ri D_k(\eta_+^k)_{\mu\nu}-\ri D_4\delta_{\mu\nu}\, ,
\label{(4.3a)}
\\[8pt]
\big(SS^\dagger\big)_{\mu\nu}&=&
-D^2\delta_{\mu\nu}+\ri (B_k+E_k)
(\eta_+^k)_{\mu\nu}\, ,
\label{(4.3b)}
\end{eqnarray}
respectively, where the $(\eta_+^i)$ are the usual 't Hooft matrices \cite{14},
$(\eta_+^i)_{jk}=\epsilon_{ijk},\,
(\eta_+^i)_{j4}=\delta_{ij}$. The important point for our considerations is that in a self-dual background the operator (\ref{(4.3b)}) is identical to the fluctuation operator
$\big[-D^2\delta_{\mu\nu}+2\ri F_{\mu\nu}]$ appearing on the left-hand side of (\ref{(1.1)}). Hence for a self-dual background the fluctuation operator
is the SUSY partner of $-D^2\I1_4$ and the formula
\begin{equation}
(\delta A_\mu)^{(\nu)}=\frac{\ri}{\omega}S_{\mu\nu}\phi,
\qquad\hbox{\small where}\quad
-D^2\phi=\omega^2\phi,
\qquad
\omega^2\neq0,
\label{(4.4)}
\end{equation}
provides us with four orthogonal fluctuation eigenmodes. (From the explicit
expressions of the 't Hooft matrices one sees that $(\delta A_\mu)^{(4)}$ is
a gauge mode, and the other three satisfy the transversality condition.)
Formula (\ref{(4.4)}) is
 known \cite{9,10}, but we thought it worth pointing out that it is a manifestation of the
 purely vectorial supersymmetry (\ref{(4.3a)})-(\ref{(4.3b)}) and that the latter is just
 the $m=\2$ special case in the series of supersymmetries (\ref{(4.2a)})-(\ref{(4.2b)}).
 Finally, we remark that for the BPS-monopole gauge potential (\ref{(2.1)}) formula (\ref{(4.4)}) produces bound vector-Higgs fluctuations with $j=\lambda,\,\lambda\pm1$ form any bound state of
(\ref{(2.5a)}) with angular momentum $\lambda=1,2,\dots$, but  only the $j=\lambda-1$ series seems to have been noticed in Ref. \cite{3}.

\section{An $\su(n/n)$ superalgebra}

Since for self-dual gauge fields the operator $S^\dagger S$ of (\ref{(4.2b)}) reduces to $-D^2\I1_n$, where $n=2(2m+1)$, it has an obvious $\su(n)$ symmetry, and we wish to show that the generators of $\Sigma_-^i$ of this $\su(n)$ induce an $\su(n/n)$ superalgebra which commutes with
$Q^2$. For this we first construct super-partner generators $\Sigma_+^i=U\Sigma_{-}^{i}U^\dagger$
(given explicitly for $\su(2)$ in (\ref{(3.4)})) which, together with the $\Sigma_-^i$ ,
 span an $\su(n)\oplus\su(n)$ Lie algebra that commutes with $Q^2$. Note that since the
 $\Sigma_{\pm}^i$  generate the defining representation of their respective $\su(n)$'s, they,
 and hence the diagonal generators $\Sigma^i=\Sigma_+^i+\Sigma_-^i$ , close (modulo central terms)
 with respect to commutation and anticommutation:
\begin{equation}
\big[\Sigma^i,\Sigma^j\big]=\ri f_{ijk}\Sigma^k,
\qquad
\big\{\Sigma^i,\Sigma^j\big\}=\ri d_{ijk}\Sigma^k
+c\delta_{ij}\I1_{2n},
\label{(5.1)}
\end{equation}
where the $f$'s and $d$'s are structure constants in a trace-orthogonal basis. We then introduce the odd (i.e. anticommuting with $\gamma_5$) Hermitian quantities
\begin{equation}
Q_1\equiv Q,\quad
Q_2\equiv \ri\gamma_5Q_1,
\quad\hbox{and}\quad
Q_\alpha^i\equiv Q_\alpha\Sigma^i,\quad
\alpha= 1,2,\quad
\gamma_5=\left(\begin{array}{cc}
\I1_n&\\&-\I1_n\end{array}\right),
\label{(5.2)}
\end{equation}
which are $\su(n)\oplus\su(n)$ scalars and vectors, respectively, and commute with $Q^2$.
Because of (\ref{(5.1)}) the anti-commutators of the operators (\ref{(5.2)}),
\begin{eqnarray}
&&\big\{Q_\alpha,Q_\beta\big\}=2\delta_{\alpha\beta}\,Q^2
\nonumber\\
&&\big\{Q_\alpha,Q_\beta^i\big\}=2Q^2\delta_{\alpha\beta}\,\Sigma^i \nonumber\\
&&\big\{Q_\alpha^i,Q_\beta^j\big\}=Q^2\,\delta_{\alpha\beta}\,\big\{\Sigma^i,\Sigma^j\big\}-
Q^2\epsilon_{\alpha\beta}\big[\Sigma^i,\Sigma^j](\ri\gamma_5),\qquad(\alpha,\ \beta=1,2)
\label{(5.3)}
\end{eqnarray}
are linear combinations of the $\Sigma^i_{\pm}$ (times $Q^2$) and a central term. Thus, for each fixed nonzero eigenvalue of $Q^2$, the  even operators $\Sigma^i_\pm$ and the odd ones (\ref{(5.2)}) generate a superalgebra \cite{15}, and since there are $2n^2$ linearly independent hermitian quantities in (\ref{(5.2)}) it is clear that this
superalgebra is $\su(n/n)$.

Note that if $S^\dagger S$ allows for an additional symmetry operator
$K$ which commutes with the $\Sigma^i_-$, then the operator $UKU^\dagger\oplus K$ will commute not only with $Q$, but
also with the whole $\su(n/n)$ superalgebra constructed above.

For the Dirac Laplacian $\D^2$ (\ref{(1.3)}) the superalgebra $\su(n/n)$ reduces to $\su(2/2)$, and it is easy to identify the  $\su(2/2)$ superalgebra found by
D'Hoker and Vinet \cite{8} as this superalgebra, in the special case when the self-dual
gauge potential is (\ref{(3.1)}). For the latter gauge potential the operator $S^\dagger S$ admits
also the Kepler symmetry discussed in Sec. 3 and its generators $\Lambda^i$ (\ref{(3.2)})
commute with $\Sigma_-^i=\sigma^i/2$. Therefore $\Lambda^i$ and $K^i$ give rise to
symmetries of $\D$ which commute with the whole $\su(2/2)$ superalgebra, as a special case of the situation
mentioned in the previous paragraph.
Thus, finally, the full symmetry algebra of $\D^2$ for the gauge-potential (\ref{(3.1)})
is $\su(2/2)\oplus\ort(4)$ for the bound states and $\su(2/2)\oplus\ort(3,1)$ for the
scattering states.

\kikezd{{\bf Acknowledgement}}. We are indebted to P. Forgacs  for useful discussions.

\kikezd{Note added.}
Neither the original text nor the list of references,
published as {\sl Int. Journ. Mod. Phys.}  {\bf A4}, 5277-5285 (1989), have been updated.
We dedicate this electronic version to the memory of
Lochlainn O'Raifeartaigh, our late  teacher, collaborator and friend.

\end{document}